\documentclass[conference]{IEEEtran}
\ifCLASSINFOpdf
\usepackage{caption}
\usepackage{epsfig}
\usepackage{color}
\usepackage{amssymb}
\usepackage{subcaption}
\begin{document}
%
\title{Search Intelligence: Deep Learning For Dominant Category Prediction}


%
\author{\IEEEauthorblockN{Zeeshan Khawar Malik, Mo Kobrosli and Peter Maas}
http://www.ebayclassifiedsgroup.com\\
\IEEEauthorblockA{Ebay Inc.
\\ Email: zmalik, mkobrosli, pmaas@ebay.com }}


\maketitle
\begin{abstract}
Deep Neural Networks, and specifically fully-connected convolutional neural networks are achieving remarkable results across a wide variety of domains. They have been trained to achieve state-of-the-art performance when applied to problems such as speech recognition, image classification, natural language processing and bioinformatics. Most of these “deep learning” models when applied to classification employ the softmax activation function for prediction and aim to minimize cross-entropy loss. In this paper, we have proposed a supervised model for dominant category prediction to improve search recall across all eBay classifieds platforms. The dominant category label for each query in the last 90 days is first calculated by summing the total number of collaborative clicks among all categories. The category having the highest number of collaborative clicks for the given query will be considered its dominant category. Second, each query is transformed to a numeric vector by mapping each unique word in the query document to a unique integer value; all padded to equal length based on the maximum document length within the pre-defined vocabulary size. A fully-connected deep convolutional neural network (CNN) is then applied for classification. The proposed model achieves very high classification accuracy compared to other state-of-the-art machine learning techniques.
\end{abstract}

\section{Introduction}

In recent years Deep Belief Networks have achieved remarkable results in natural language processing \cite{kim2014convolutional}, computer vision \cite{hinton2012improving}\cite{malik2016multilayered} and speech recognition \cite{hinton2012improving} tasks. Specifically, within natural language processing, modeling information in search queries and documents has been a long-standing research topic \cite{deerwester1990indexing}\cite{gao2004dependence}. Most of the work with deep learning has involved learning word vector representations through neural language models \cite{le2014distributed}\cite{malik2014novel}\cite{onlinegenlap} and performing composition over the learned word vectors for classification \cite{collobert2011natural}. 

The optimal transformation in our case was to map each query document to a single numeric vector by assigning a single numeric value to each unique word across all query documents. A second phase was then employed by mapping the numerically transformed query vectors to a random embedding space having a uniform distribution between -1 and 1. This helped far more in reducing the distance between queries having similar words while further discriminating queries far on the data space having more dissimilar words. Another suitable criteria that is applicable to our problem is proposed by Johnson and Zhang \cite{johnson2014effective} in 2014, where they propose a similar model, but swapped in high dimensional `one-hot' vector representations of words as CNN inputs. 

Convolutional Neural Networks (CNN) are biologically-inspired variants of Multiple Layer Perceptrons (MLP). They utilize layers with convolving filters that are applied to local features \cite{lecun1998gradient} originally invented for computer vision. Convolutional neural networks have also been shown to be highly effective for natural language processing and have achieved excellent results in information retrieval \cite{shen2014learning}, semantic parsing \cite{yih2011learning}, sentence modeling \cite{kalchbrenner2014convolutional} and other traditional natural language processing tasks \cite{collobert2011natural}.

Before going into the details of our model architecture and results, we will first narrate the work we did to prepare our query data for modelling.

\section{Query Data Preparation}

The advertisements in eBay's classifieds platforms are classified according to a pre-defined hierarchy. The first level (L1) of this hierarchy categorizes advertisements into general groupings like `buy \& sell', `cars \& vehicles', `real state', `pets', `jobs', `services', `vacation rentals' and `community'. The second level (L2) further classifies each L1-category with many subclasses with more specificity. The third level (L3) further classifies and so on. Most platforms terminate the hierarchy at a level of three or four. In this paper we will only demonstrate the results of our work related to L1-category query classification. 

For each keyword search initiated within a user session at the all-advertisement level (all-advertisement level means a search across all inventory with no category restrictions employed), the chain of actions on that search is analysed. When that sequence of actions results in a view of an advertisement within a specific category, that particular category is scored with a dominance point for the given query. There are many noisy factors that must be accounted for when applying this technique. Among them include factors like bots, redundant query actions, filtering out conversions to categories that no longer exist and filtering out queries without enough conversions. 


The dominance of category for each query document in the last 90 days is computed on the basis of the maximum number of collaborative clicks for each L1-category. The category with the highest number of clicks is considered the dominant category for that query. This also enabled us to produce the first highest, second highest and third highest dominant category and their respective conversion rates for each query. The conversion rate per query is calculated by counting the total number of clicks for each category divided by the total number of clicks for that query. 

Finally all query documents for the last 90 days are standardized by transforming them to lower-case, removing duplicate queries, extra spaces, punctuations and all other noise factors. A single pattern from each L1-category of the final preprocessed data ready to be used for learning is shown in Table \ref{preprocessed}.

In Table \ref{preprocessed} the CategoryID feature is used as a label for supervised learning using a deep convolutional neural network. The total distinct query patterns for most of the categories in the last 90 days ranges between 5000 to 7000.

\begin{table*}[ht]
\centering
\caption{A Single Unique Pre-processed Pattern From each L1-Category}
\begin{tabular}{|c|c|c|c|c|}
\hline
Category Name & CategoryID & Query & Category Conversion-Rate & Total Patterns  \\
\hline
cars \& vehicles&27 & 2007 civic & 0.9857 \ 98\% & 5000 - 7000 \\
\hline
jobs & 45   & cash jobs & 0.7051 \ 70\% & 5000 - 7000  \\
\hline
services & 72 & makeup artist & 0.8911 \ 89\%  & 5000 - 7000 \\
\hline
buy \& sell & 10 & air conditioner & 0.9783 \ 97\% & 5000 - 7000 \\
\hline
vacation rentals & 800 & sherkston shore & 0.4694 \ 46\% & 2000 - 3000\\
\hline
pets & 112 & western saddle & 0.8268 \ 82\% & 5000 - 7000 \\
\hline
real state& 34& mortgage & 0.4782 \ 47\% & 5000 - 7000 \\
\hline
community& 1& christmas markets & 1 \ 100\% & 2000 - 3000 \\
\hline
\end{tabular}
\label{preprocessed}
\end{table*}

\section{Model Architecture}

\begin{figure*}[ht]
\centering
\includegraphics[width=18cm, height=4 in]{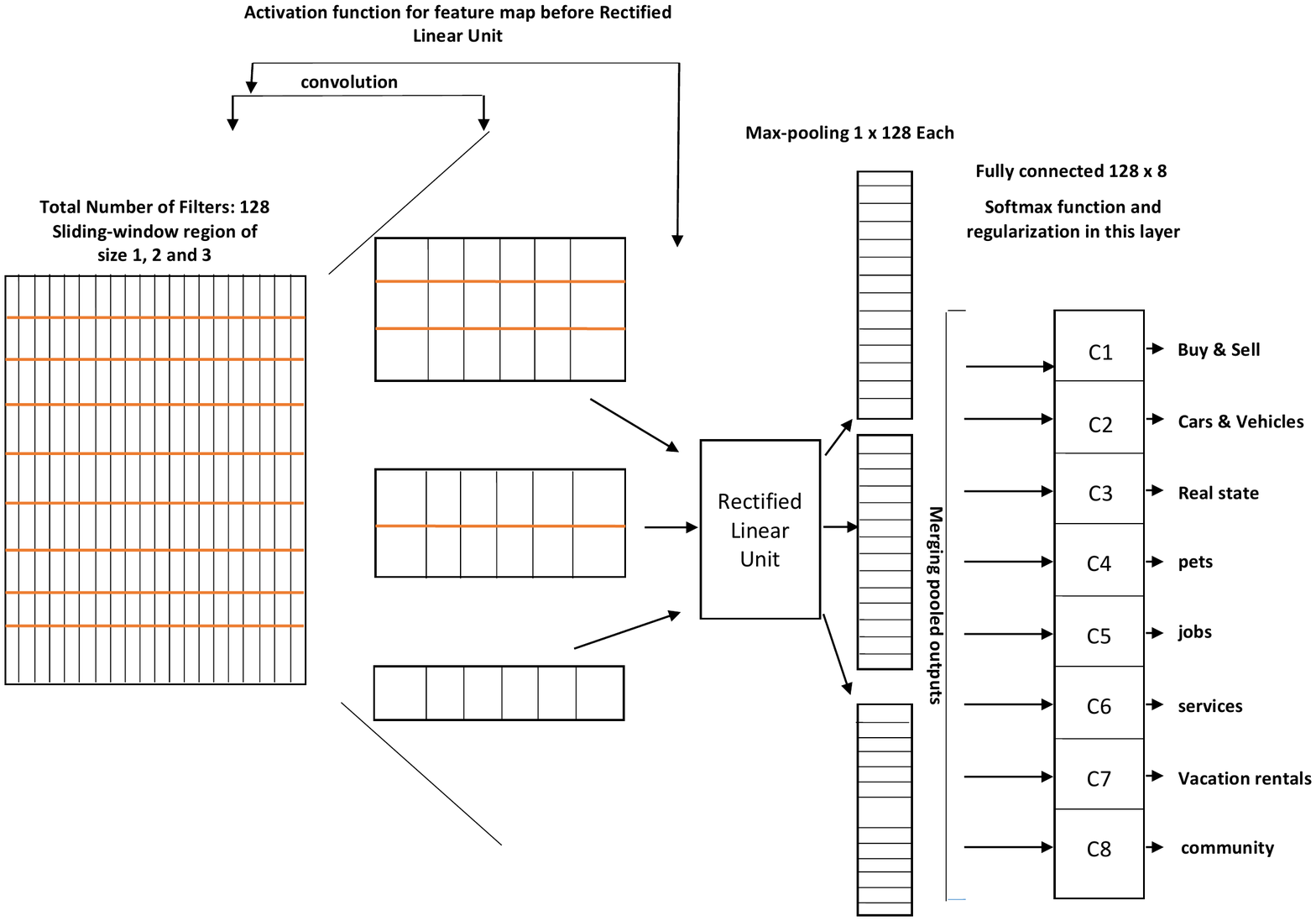}
\caption{Model Architecture}
\label{Architecture}
\end{figure*}

The model architecture shown in Figure \ref{Architecture}, follows \cite{zhang2015sensitivity} and \cite{kim2014convolutional}. Let ${\bf x}_{i} \ \epsilon \ \mathbb{R}^{k}$  be the $k$ dimensional transformed numeric vector for each query document mapping each word in the query document to an integer within the defined vocabulary size. 

Suppose we have a given query document $D = (w_{1}, w_{2}, ..., w_{N})$ with vocabulary $V$. CNN requires vector representation of data that can uniquely preserve internal locations (word order in this case) as input. The chosen best and straight forward representation would be to treat each word	as a pixel, treat $D$ as if it were an image of $|D| \times 1$ pixel, and to represent each pixel (i.e. each word) with a unique numeric value. As a running real-time example  suppose that query document D = \{"giving", "away", "free", "free"\} and we associate the words with unique numeric value. Then we have the document vector as:-
\begin{equation}
{\bf x} = [1235, 1643, 1245, 1245, 0, 0, 0, 0, 0]
\end{equation}


All the query document vectors are padded to equal length based on the maximum document size of the last nighty days query corpus is represented as

\begin{equation}
{\bf x}_{1:n} = {\bf x}_{1} \oplus {\bf x}_{2} \oplus {\bf x}_{3} \oplus ... {\bf x}_{n}
\end{equation}


where $\oplus$ is the concatenation operator. Let ${\bf x}_{i:i+j}$ refer to the concatenation of words ${\bf x}_{i}, {\bf x}_{i+1},...,{\bf x}_{i+j}$ of a single query document with the unique numeric conversion for each word. The filtration of ${\bf w} \ \epsilon \ \mathbb{R}^{hk}$ is considered as a convolution operation, which is applied to a window of $h$ words to produce a new feature. Supposedly, a feature $c_{i}$ is generated from a window of words ${\bf x}_{i+h-1}$ by

\begin{equation}
c_{i} = f({\bf w}\cdot{\bf x}_{i:i+h-1} + b)
\end{equation}     

\captionsetup{justification=centering,margin=2cm}

where $b \ \epsilon \ \mathbb{R}$ is a bias term and $f$ is a non-linear activation function such as the tangent hyperbolic function. The filter is applied to each possible window of words throughout the whole set ${\{\bf x}_{1:h}, {\bf x}_{2:h+1}, ..., {\bf x}_{n-h+1:n}\}$ to produce a feature map. 

\begin{equation}
{\bf c} = [c_{1}, c_{2}, ..., c_{n-h+1}]
\end{equation}

with ${\bf c} \ \epsilon \ \mathbb{R}^{n-h+1}$. The feature mapping is followed by the rectified linear unit which zeros out negative values and produces sparse activations. Next comes the max-pooling layer which captures the most significant feature, the one with the highest value for each feature map. 

Above, we explained the process by which one feature is extracted from one filter. The proposed model uses multiple filters with varying window sizes to obtain multiple features. These extracted significant features form the penultimate layer and are passed to a fully connected softmax layer whose output is the probability distribution over 8 labels.

We have employed dropout for regularization on the penultimate layer \cite{hinton2012improving}. Dropout helps in preventing co-adaptation of hidden units by randomly dropping with a certain probability. Given the penultimate layer $z=[\hat{c}_{1}, ....,\hat{c}_{m}]$, dropout uses

\begin{equation}
y = {\bf w} \cdot ({\bf z} \circ {\bf r} + b),
\end{equation}

where $\circ$ is the element-wise multiplication operator and ${\bf r} \ \epsilon \ \mathbb{R}^{m}$ is a `masking' vector of Bernoulli based random variables with probability $p$ being 1. In this way the dropout mechanism for regularization on the penultimate layer stochastically disables a fraction of its neurons. This ultimately prevent neurons from co-adapting and forces them to learn individually useful features. The fraction of neurons to keep enabled is defined by the dropout keep probability input to the network. 

Table \ref{configuration} summarizes the configuration details of the employed deep convolutional neural network which significantly solved the dominant category prediction problem across several eBay Classifieds platforms.  The first column defines the length of embedding layer size which maps the input to an embedding space. The filter size narrates the number of words we need to consider in each convolutional filter.  The total number of filters for each window of size 1, 2 and 3 are 128. The batch size and number of epochs for training are set to 64 and 100. The maximum length of query sequence in our case is 10 and the total number of L1-Category classes are 8. The training time of the algorithm for the 90 days of data is approximately 50 minutes on an Intel Core i7 with 2.8 GHz specification. 

The summary statistics of our pre-computed dominant category prediction dataset are shown in Table \ref{summary_statistics} which describes the total number of classes, average sentence length, vocabulary size, training and testing data length.

\begin{table*}[ht]
\centering
\caption{Configuration of Deep Convolutional Neural Network For L1 Dominant Category Prediction}
\begin{tabular}{|c|c|c|c|c|c|c|c|}
\hline
 Embedding Layer Dim. & Filter Sizes &Number of Filters & Dropout Keep Probability & Batch Size & No. of Epochs & Sequence Length & No. of Classes  \\
\hline
128&1, 2, 3 & 128 &0.5 & 64 & 100 & 10 & 8\\
\hline
\end{tabular}
\label{configuration}
\end{table*}

\begin{figure*}
\centering
\begin{subfigure}{.37\textwidth}
  \centering
  \includegraphics[width=3.0 in, height=2 in]{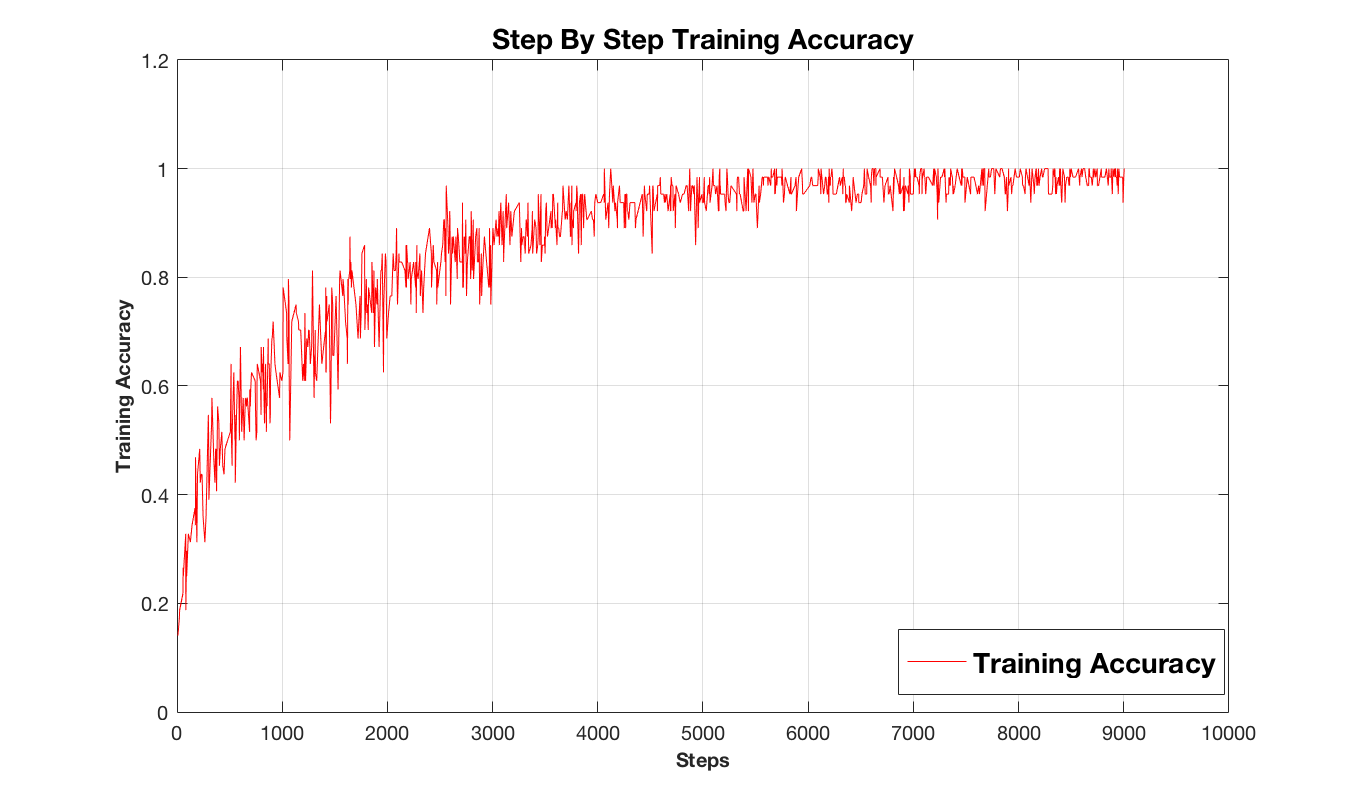}
  \caption{Training Accuracy}
  \label{fig:sub1}
\end{subfigure}
\begin{subfigure}{.37\textwidth}
  \centering
  \includegraphics[width=3.0 in, height=2 in]{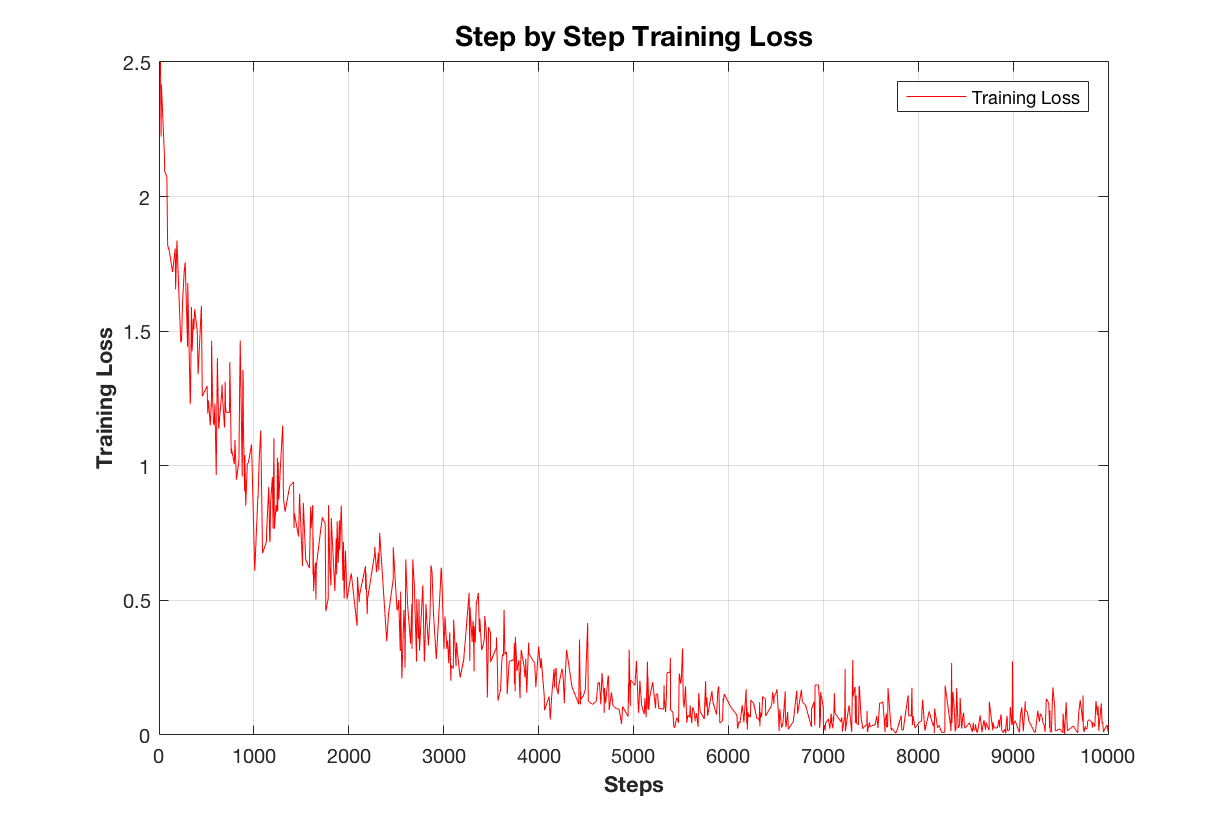}
  \caption{Training Loss}
  \label{fig:sub2}
\end{subfigure}
\centering
\caption{Training Accuracy \& Loss of CNN}
\label{fig:test}
\end{figure*}

\begin{table*}[ht]
\centering
\caption{Summary Statistics of the Dataset}
\begin{tabular}{|c|c|c|c|c|c|}
\hline
 Data & Number of Classes & Average Sentence Length & Vocabulary Size & Training Size & Testing Size  \\
\hline
Dominant-Category & 8 & 2 & 12812 & 32088 & 32087\\
\hline
\end{tabular}
\label{summary_statistics}
\end{table*}

\begin{table*}[ht]
\centering
\caption{Results of the proposed well-tuned CNN model against other methods}
\begin{tabular}{|c|c|c|c|c|c|}
\hline
 Model Type & Number of Days & Training Date Range  & Testing Date Range & Training Accuracy& Testing Accuracy   \\
\hline
CNN {\bf (Proposed)} & Past 90 Days & 28-06-2016 to 28-09-2016 & 07-06-2016 to 07-09-2016 & {\bf 0.9999} {\bf 99.9 \%} & {\bf 0.98529} {\bf 98.5 \%} \\
\hline
CNN {\bf (Proposed)} & Past 90 Days & 28-06-2016 to 28-09-2016 & 28-02-2016 to 28-05-2016 & {\bf 0.9999 } {\bf 99.9 \%}& {\bf 0.95891} {\bf 95.8 \%}  \\             
\hline
\hline
CNN-static \cite{kim2014convolutional} & Past 90 Days & 28-06-2016 to 28-09-2016 & 07-06-2016 to 07-09-2016 &  {\bf 0.9857 } {\bf 98.5 \%}& {\bf 0.95926} {\bf 95.9 \%}  \\             
\hline
CNN-static \cite{kim2014convolutional} & Past 90 Days & 28-06-2016 to 28-09-2016 & 28-02-2016 to 28-05-2016 & {\bf 0.9737 } {\bf 97.37 \%}& {\bf 0.9317} {\bf 93.17 \%}  \\             
\hline
\hline
CNN-non-static \cite{kim2014convolutional} & Past 90 Days & 28-06-2016 to 28-09-2016 & 28-02-2016 to 28-05-2016 & {\bf 0.9999 } {\bf 99.9 \%}& {\bf 0.95891} {\bf 95.8 \%}  \\             
\hline
CNN-non-static \cite{kim2014convolutional} & Past 90 Days & 28-06-2016 to 28-09-2016 & 28-02-2016 to 28-05-2016 & {\bf 0.9953 } {\bf 99.53 \%}& {\bf 0.94043} {\bf 94.043 \%}  \\             
\hline
\hline
MLP with two hidden layers & Past 90 Days & 28-06-2016 to 28-09-2016 & 07-06-2016 to 07-09-2016 & 0.563486 56.35 \%  &  0.559056  55.91 \% \\
\hline
MLP with two hidden layers & Past 90 Days & 28-06-2016 to 28-09-2016 & 28-02-2016 to 28-05-2016& 0.563486  56.35 \%  &  0.549894  54.98 \% \\
\hline
\hline
MLP with single hidden layer & Past 90 Days & 28-06-2016 to 28-09-2016 & 07-06-2016 to 07-09-2016 & 0.483046  48.31 \%  &  0.479556  47.95 \% \\
\hline
MLP with single hidden layer & Past 90 Days & 28-06-2016 to 28-09-2016 & 28-02-2016 to 28-05-2016& 0.483046  48.31 \%  &  0.483915  48.39 \% \\
\hline
\hline
LSTM RNN Network & Past 90 Days & 28-06-2016 to 28-09-2016 & 07-06-2016 to 07-09-2016 & 0.658262 65.83 \% & 0.651895 65.19 \% \\
\hline
LSTM RNN Network & Past 90 Days & 28-06-2016 to 28-09-2016 & 28-07-2016 to 28-04-2016 & 0.658262 65.82 \% & 0.630651 63.06 \% \\
\hline
\hline
LSTM Bi-RNN Network & Past 90 Days & 28-06-2016 to 28-09-2016 & 07-06-2016 to 07-09-2016 & 0.536496 53.65 \% & 0.529887 52.98 \% \\
\hline
LSTM Bi-RNN Network & Past 90 Days & 28-06-2016 to 28-09-2016 & 28-07-2016 to 28-04-2016 & 0.536496 53.65 \% & 0.505335 50.05 \% \\
\hline
\hline
\end{tabular}
\label{model_result}
\end{table*}

\section{Results \& Discussion}

Results of the proposed model for the dominant category prediction problem compared to other state-of-the-art methods are listed in Table \ref{model_result}. The proposed well-tuned deep convolutional neural network simply outperformed its variations and other models. We tested the predictive accuracy by first using few days different testing data from training shown in the first row and fourth column of Table \ref{model_result} for every model type. The CNN model produced a very high training and testing accuracy of 99.9 \% and 98.5 \%. Secondly we tried testing completely different days testing data from training and the resulting outcomes are shown in the second row of Table \ref{model_result} for every model type. This is our worst case scenario where we have used a completely different testing data for dominant category prediction but still the CNN model has produced a very high testing accuracy of 95.8 \%. The major advantage with CNN compared to other state-of-the-art approaches is its added capability to learn invariant features. This capability of CNN to make the convolution process invariant to translation, rotation and shifting helps in approximating to the same class even when there is a slight change in the input query document. 

The step by step training accuracy and loss of our convolutional neural network model are also shown in Figure \ref{fig:sub1} and \ref{fig:sub2}.  Initially the accuracy was noted very low but gradually it improved at each training step and almost reached to one in the end as shown in Figure \ref{fig:sub1}. Similarly, the loss was very high in the beginning, but almost reached to zero in the end as shown in Figure \ref{fig:sub2}. This clearly shows the convergence of the proposed well-tuned deep convolutional neural network.

The multiple layer perceptron model with an empirically evaluated one and two hidden layers of size 200 did not perform effectively well and produced a predictive accuracy of 55.91 \%  and 54.98 \% on both of the testing sets. We also further tried to increase the count of hidden layers to explicitly add the certain level of non-linearity but still the predictive accuracy more or less remained constant. Furthermore we tried running Long Short Term Memory (LSTM) recurrent neural networks which are shown to outperform other recurrent neural network algorithms specifically for language modelling \cite{sundermeyer2012lstm}. However, in our case there is no sequence to sequence connection between the current and previous activations of the sequential query patterns, the maximum predictive accuracy that LSTM recurrent neural network could produce was 63.06 \% and 65.19 \% for both the testing datasets. The Bi-directional recurrent neural network worked a little worse compared to LSTM network and produced a predictive accuracy of 52.98 \% and 50.05 \% on both the testing datasets.

%

\section{Conclusion}

In the present work we have described a tuned, fully connected CNN that outperformed its variants and other state-of-the art ML techniques. Specifically, in query to category classification across several eBay Classifieds platforms. Our results integrate to evidence that numeric vector mapping to random uniformly distributed embedding spaces proves more suitable both computationally and performance wise in comparison to word2vec. Specifically for datasets having a limited vocabulary corpus (between 10,000 to 15,000 words) and few words (between 2 to 3) in each query document.

\section{Acknowledgement}

The first and second authors are grateful to Johann Schweyer for his contribution in query normalization and aggregation. We are also extremely thankful to Brent Mclean VP, CTO, eBay Classifieds for his kind support and encouragement throughout this dominant category prediction project.


\end{document}